# Trustworthy Smart Fabs via Professional Proxies: Scaling Safe and Sustainable by Design (SSbD) through Industrial Data Spaces

This work was accepted for presentation at the 32nd IEEE ICE/ITMC Conference, Porto, Portugal, 2026 but was subsequently withdrawn prior to publication due to submission volume limits. It is currently under consideration for publication elsewhere.

Han-Teng Liao*
*Independent Researcher*
Penang, Malaysia
h.liao@ieee.org
0000-0003-1081-5599

Chang-Yi Kao
*Dept. Computer Science and Information Management*
*Soochow University*
Taipei, Taiwan
edenkao@scu.edu.tw
0000-0003-0075-0787

Karen Ang
*Independent Researcher*
Kulim, Malaysia
karenang118@gmail.com
0009-0008-5923-0106

***Abstract*****—** The convergence of the 2026 European Union Safe and Sustainable by Design (SSbD) framework, Corporate Sustainability Due Diligence Directive (CSDDD), and Carbon Border Adjustment Mechanism (CBAM) introduce a severe governance bottleneck for advanced semiconductor manufacturing facilities ("Smart Fabs"). Regulatory compliance demands have surpassed the capacity of manual corporate reporting, creating a direct conflict between multi-stakeholder transparency and corporate data privacy. This paper addresses this challenge by introducing a zero-trust socio-technical orchestration framework that operationalizes a six-layer SSbD reference architecture within trustworthy industrial data spaces. We propose a shift from reactive automation to autonomous governance through "Professional Proxies"—role-based agentic workflows executing within hardware-isolated trust zones. Structured as an interoperable network protocol stack, the framework coordinates an automated, five-step "relay race" between Facility, Process Engineering, and Finance proxy teams to align factory-floor yield models with macro-level sustainability mandates. By executing Virtual Metrology (VM) predictions and Federated Machine Learning (FML) inside hardware-rooted Trusted Execution Environments (TEEs), this architecture resolves the Data Sovereignty Paradox, demonstrating how fabs can export cryptographically signed compliance tokens via International Data Spaces (IDS) connectors without exposing proprietary process recipes. Ultimately, this framework provides technology managers with a verifiable, evidence-based pathway toward resilient, net-zero Industry 5.0 ecosystems.

***Keywords—Agentic Workflows, Trusted Execution Environments, Socio-Technical Systems, Industry 5.0, RegTech***

## I. Introduction: From Automation to Agency

### A. Context: Geopolitical Reconfiguration of Value Chains

The rapid reconfiguration of global value chains (GVCs)—driven by geopolitical tensions and the strategic deployment of the EU and US Chips Acts—has placed advanced semiconductor manufacturing facilities ("Smart Fabs") in a precarious operational position. Recent disputes regarding advanced artificial intelligence (AI) logic chips and critical chemical supply chains have caused disruptions that ripple across the automotive, energy, and defense sectors [1]. Supply chain safety and structural resilience have thus risen to the level of national security priorities, necessitating integrated cyber-physical frameworks that can coordinate real-time operational decisions across pricing, demand forecasting, and capital expenditure to maintain an agile, compliant, and sustainable value chain [2].

In this environment, modern fabrication facilities are no longer merely centralized manufacturing hubs; they have evolved into massive data engines subject to overlapping, cross-border international mandates. The immediate challenge posed by the Corporate Sustainability Due Diligence Directive (CSDDD) and the March 2026 European Union Safe and Sustainable by Design (SSbD) guidelines [3], [4] is that regulatory compliance has surpassed the threshold of manual human oversight or periodic retroactive auditing. Consequently, the digital transformation of Environmental, Social, and Governance (ESG) requirements must now scale dynamically across companies, sectors, and jurisdictions, necessitating a fundamental architectural transition from static factory automation to active, autonomous regulatory agency.

### B. Motivation: Redefining Autonomy as Regulatory Agency

In high-stakes, capital-intensive manufacturing environments, true autonomy must be redefined as active regulatory agency within the domain of Regulatory Technologies (RegTech). To establish and maintain trustworthy industrial relationships among global supply chain stakeholders, autonomous systems must function as Professional Proxies—specialized agentic workflows acting on behalf of distinct corporate and engineering practices:

- **The Fab Facility Manager Proxy:** Autonomous agentic workflows that monitor, verify, and manage chemical configurations to ensure they remain safe and sustainable by design.
- **The Process Engineering Manager Proxy:** Agentic workflows that reconcile strict data integrity, privacy and security requirements within the Industrial Internet of Things (IIoT)..
- **The Fab Procurement & Finance Accountant Proxy (Digital Auditing & Financial Integrity):** Agents that autonomously calculate and verify Embedded Carbon Emissions and Product Carbon Footprints (PCF) for Carbon Border Adjustment Mechanism (CBAM) compliance, presenting immutable proofs to external clearinghouses without exposing sensitive factory-floor trade secrets or proprietary equipment telemetry.

### C. Technology: From Physical Metrology to Virtual Fabs

Sub-5 nm scaling has rendered traditional physical metrology and Statistical Process Control (SPC) insufficient for real-time monitoring. This necessitates a transition to Virtual Metrology (VM), where machine learning predicts product quality and mitigates chamber variations directly from equipment status data [5], [6]. By integrating deep learning and novelty detection, Smart Fabs achieve precision and fault detection beyond traditional limits [7], [8]. The "Virtual Fab" may transform manufacturing from reactive quality checks into proactive, real-time quantification of both yield and environmental footprints.

Operationalizing this trust requires grounding agentic logic in recognized international metrology benchmarks, specifically the ITU-T L-series emission standards. Because these ITU standards are increasingly aligned with the Science Based Targets initiative (SBTi)—which provides the rigorous tools and guidance for companies to set greenhouse gas (GHG) reduction targets in line with a 1.5°C climate trajectory—they serve as the definitive "scientific truth" for industrial agents. By embedding SBTi-compliant trajectories directly into the Professional Proxy's core code, the system moves from aspirational corporate ESG goals to technically verifiable outcomes. This data-driven framework establishes the structural foundation for building robust innovation and business ecosystems, operationalizing the advanced helix model of collaborative practices for ecosystem innovation [9].

### D. Objectives: Architecting Data-Driven Evidence-Based Trust in Industry 5.0

The overarching objective of this study is to bridge the structural divide between real-time manufacturing optimization and macro-level sustainability mandates within the semiconductor industry. Rather than treating compliance as an out-of-band, retrospective accounting burden, this research develops an inline, automated engineering architecture. Consequently, the central research question of this study is formulated as follows:

How can autonomous agents be engineered as trustworthy professional proxies to operationalize a zero-trust compliance fabric that resolves the structural friction between local factory autonomy, global value chain reconfigurations, and shifting international industrial standard-making?

The primary bottleneck preventing the execution of this agentic vision is the Data Sovereignty Paradox: the systemic conflict of how a firm can share granular, verified compliance data across an adversarial value chain without losing control of its core intellectual property or manufacturing recipes. By designing a multi-agent framework rooted in hardware-isolated cryptographic trust zones and integrating International Data Spaces (IDS) Connectors across the factory perimeter [10], and embedding declarative Policy-as-Code engines via Open Policy Agent (OPA) [11], [12], this paper provides the actionable framework necessary to codify multi-stakeholder trust through three strategic pillars:

- **Chemical Transparency (Environmental):** Utilizing decentralized Digital Product Passports (DPP) to verify chemical toxicity, provenance, and circularity metrics across global chemical vendors.
- **Socio-Technical Resilience (Privacy/Labor):** Synchronizing technical factory-floor controls with organizational governance processes to ensure local industrial autonomy remains compliant with global human rights mandates and institutional labour regimes.
- **Financial Integrity (Audit/Finance):** Linking supply chain and trade financing directly to verifiable sustainability metadata, thereby preventing corporate greenwashing and ensuring the structural validity of ESG-linked credit lines for energy-intensive components.

## II. Existing Designs & Previous Work

### A. Sector Baseline and Sustainability Gaps

The IEEE International Roadmap for Devices and Systems (IRDS™) chapter on Environment, safety, health & sustainability (ESHS): environmental sustainability of the semiconductor facilities (ESSF) establishes the benchmark for facility sustainability, some with specific Key Performance Indicators (KPIs):

- **Water Evaporative Losses:** Tracked via KPI-1 (Equation #1) to evaluate water loss from evaporative cooling.
- **Freshwater Usage:** Tracked via KPI-2 (Equation #2) to measure freshwater usage.
- **Reuse Efficiency:** Governed by KPI-3 (Equation #3) to measure reuse efficiency under the Total Alternative Water Optimization (TACO) framework
- **System Discharges:** Additional metrics covering non-beneficial Ultra-Pure Water (UPW) usage and hazardous waste generation.
- **Energy Consumption Impact:** Utilized SEMI S23 Equivalent Energy Consumption Factors (ECF).
- **Financial Benchmarking:** Quantified via Cost of Ownership (CoO) (Equation #4) and normalized operational equations (#5–#7).

According to the IRDS ESHS ESSF, though digital twin models of both water and energy are built to capture environmental impacts, several critical gaps remain unaddressed. No mathematical formulas for Greenhouse Gasses (GHG) and real-time chemical usage intensity were used:

- *GHG Scope 1 Abatement:* Fabs remain major sources; near-term mandates require pushing tool-level abatement efficiencies from an 80% baseline up to 95%-99%.
- *Scope 3 Footprint Blindspots:* Current frameworks lack the data granularity to track the embodied energy of supply chain process materials (chemicals and gases), requiring integration of in-situ and on-site recycling.
- *Narrative vs. Executable Goals:* Critical parameters like hazardous chemistry minimization and per- and polyfluoroalkyl substances (PFAS) elimination are treated as narrative design goals rather than active, tool-level operational constraints.

### B. Supply Chain, ESG Ecosystems, and Industrial Data Spaces

The transition toward agentic governance in the semiconductor industry builds upon established digital platforms for responsible supply chain management, most notably TSMC's Supply Online 360 [13]. This platform serves as a milestone for digitized sustainability by integrating a Supplier Sustainability Management Module, the TSMC Supplier Sustainability Academy, and a public Supply Chain Worker Grievance Channel. By tracking closed-loop improvement measures across a common-good model, the platform manages economic, environmental, and social risks.

However, shifting 2026 SSbD and CSDDD mandates demand a granularity and velocity of Scope 3 PCF and labor data that threaten to exceed traditional portal-based management. This research identifies Supply Online 360 as the foundational infrastructure for the next stage of industrial evolution: the integration of Professional Proxies. By utilizing the platform's modules as a knowledge base and telemetry source, autonomous agents can be deployed as Digital Auditors and Privacy Stewards, transforming static reporting into a self-correcting, agentic ecosystem capable of meeting high-velocity compliance demands.

### *C. Trustworthy and Secure AI Foundations*

The foundation of a professional proxy rests on three dimensions of trust: data integrity, identity sovereignty, and infrastructure resilience.

- **Data Integrity via Synthetic Fidelity:** Achieving regulatory transparency for SSbD audits requires accurate, non-identifiable data. Generative models can produce high-fidelity, privacy-preserving synthetic records that maintain statistical utility while satisfying stringent privacy metrics [14]. This allows proxies to execute "dry run" audits or yield optimizations without exposing a fab's raw, proprietary manufacturing recipes, ensuring intellectual property remains secured while upholding global transparency mandates..
- **Identity Sovereignty via Trust Negotiation:** The legitimacy of an autonomous agent depends on its ability to incrementally establish its professional mandate. Trust negotiation frameworks provide the mechanism for digital entities to exchange credentials and establish decentralized, multi-stakeholder trust [15], such as Trusted Execution Environment (TEE) [16]. By utilizing a sovereign identity layer, proxies can represent professional groups—such as auditors or sustainability stewards—to navigate cross-border regulatory regimes safely.
- **Infrastructure Resilience through Integrated Security:** Trustworthy agency requires an infrastructure that protects sensitive industrial data at rest, in transit, and in use. Holistic cloud security paradigms advocate for integrating hardware-rooted trust to protect sensitive industrial workloads [17]. This resilience layer ensures that professional proxies can execute performance-based carbon accounting on encrypted datasets, resolving the Data Sovereignty Paradox by securing raw telemetry from exposure.

### *D. Regulatory Evolution: National Acts and Policy-as-Code*

As the first region-wide due diligence regulation, the EU Corporate Sustainability Due Diligence Directive (CSDDD) unifies fragmented national mandates, building upon the German Supply Chain Act and the Norwegian Transparency Act [18], [19]. For manufacturing and procurement organizations, the CSDDD introduces high-stakes implementation challenges across four critical stages: exploration, installation, initial implementation, and full implementation [20]. The primary hurdles involve managing the immense dynamic complexity of global supply chains while ensuring that sustainability measures are integrated cost-efficiently without compromising competitiveness.

This regulatory pressure forces a move toward RegTech and Policy-as-Code, where identifying and mitigating adverse impacts—such as those governed by the Conflict Minerals Regulation and forthcoming Forced Labour Regulation—becomes a real-time requirement. Mandatory due diligence requires a verifiable, standardized scope managed via Policy-as-Code agentic workflows, demanding meaningful human-centric orchestration in corresponding RegTech.

This digital transformation of due diligence profoundly impacts trade and production globalization. The upcoming CBAM intensifies this pressure by shifting from governance-based reporting to performance-based carbon accounting at the border, potentially impacting the socio-economic stability of emerging and developing economies (EMDEs) reliant on EU trade [21]. To mitigate regressive impacts and ensure carbon information integrity, there is a clear trajectory toward adopting ITU standards (e.g., ITU-T L.1470/L.1480) and interoperable carbon management platforms [22]. Within this context, compliance shifts from a periodic manual audit to an operational execution challenge, creating a critical role for autonomous agents engineered as trustworthy professional proxies to navigate the conflict between local industrial autonomy and global value chain mandates.

### *E. Digital Transformation Partners and Capabilities*

Previous work on ESG capability building emphasize the role of cross-disciplinary collaboration and partner mapping as essential pedagogical and organizational tools [23]. These design considerations and highlight high-tech manufacturing sustainability as a socio-technical collaborative challenges requiring "digital-ready" talents who can navigate complex stakeholder ecosystems.

While such work correctly identified the need to bridge talent training connecting engineering, accounting, auditing and content communication to ensure carbon information integrity, as evidenced by job requirements at advanced semiconductor fabs—such as TSMC and Infineon— ranging from sustainability strategists, data analysts, and content specialists, it did not foresee the emergence of agentic artifacts, and its impact on human-centric "capability building."

The rapid scaling of ESG data (e.g., real-time Scope 3 monitoring) and the high energy-intensity of AI-driven manufacturing have rendered manual partner mapping insufficient. The shift from mere "educational design" to "operational execution" necessitates exploring autonomous agents as professional proxies (likely in the forms of *AI Skill artifacts*). These agents must evolve beyond information exchange to function as "trustworthy intermediaries" that execute *Policy-as-Code* guidelines, ensuring local industrial autonomy remains compliant with global regulatory regimes.

## III. Method

To validate the feasibility, trustworthiness, and scalability of the proposed agentic architecture, this study adopts a multi-methodological engineering design research [24] framework. Building upon the physical baselines established by the IEEE IRDS™ ESHS ESSF roadmap, the methodology develops a conceptual system-of-systems (SoS). The research **architects a structural blueprint** for localized, role-based autonomous agents interacting inside hardware-isolated environments. This computational setup is evaluated against real-world regulatory constraints, including CBAM, CSDDD, and SSbD guidelines. Through this design-science approach [25], [26], we **conceptually evaluate** how effectively this SoS of trustworthy industrial data spaces can scaling safe and sustainable semiconductor manufacturing.

### A. Presented Study and Research Questions

The study evaluates the structural logic and architectural viability of a zero-trust multi-agent system. To investigate, the overarching research question consists of three operationalized research questions:

- $RQ_1$ **Operations & Autonomy**: *How can autonomous agents engineered as Process Engineering and Fab Facility Manager proxies intercept edge-level tool telemetry to execute inline SSbD compliance loops without degrading factory autonomy or wafer yield*? This is evaluated by architecting **a hardware-isolated system-of-systems structure where physical and soft-sensor data streams are secured and processed in TEE enclaves**, preventing data tempering.
- $RQ_2$ **Finance & Accountability:** *How can a Fab Procurement & Finance Accountant proxy dynamically translate physical resource metrics into verifiable accounting records to resolve various financial and regulatory tensions?* This is evaluated by designing **an inline RegTech layer that structures agentic proxy-ready schemas to use facility metrology data streams as immutable compliance proofs**, without exposing proprietary, tool-level process recipes.
- $RQ_3$ **Global Value Chain Harmonization:** *How can these localized agentic proxies interact in an open, interoperable protocol stack capable of harmonizing conflicting manufacturing regulatory demands with standard specifications to secure a resilient, net-zero Industry 5.0 ecosystem?* This question is addressed by **technology roadmapping for federated data sharing and federated learning** of related documents and specifications.

The proposed agentic architecture should allow professional proxies to share distinct analytical frameworks to solve the above operational questions.

### B. Applied Research Methods

By applying a Design Science Research (DSR) paradigm common in information systems research, this article generates a conceptual reference architecture for autonomous regulatory agency. DSR balances the dual activities of building innovative socio-technical artifacts and evaluating their structural performance within bounded industrial domains. This study treats advanced semiconductor manufacturing as a Socio-Technical System (STS), where digital agentic networks serve as the "nerves" of information, communication, and control.

Emulating the structural logic of the classic Open Systems Interconnection (OSI) model [27], the paper adopts and significantly extends the baseline 6-layer Safe and Sustainable by Design (SSbD) architecture [28] to contextualize how three professional proxy agents work conceptually to resolve the Data Sovereignty Paradox. Thus, the modular, interoperable layers or stacks should operationalize (and hopefully standardize in the future) how compliance data moves from physical edge telemetry up to cross-border value chains.

### C. Data Collection and Sources

The data collection protocol relies on secondary document analysis, with the purpose to synthesize industry consensus and ground truths with technology roadmapping while establishing the operational parameters that define the boundaries of the Data Sovereignty Paradox:

1. **Industry Roadmap and Regulatory Data:** Technical specifications, environmental thresholds, and resource targets are gathered from the IRDS roadmaps and leading foundry reports (e.g., *TSMC Sustainability Reports*), with legal compliance of the CBAM and EU SSbD mandates.
2. **Architectural and Infrastructure Specifications**: Relevant technical models such as the *International Data Spaces (IDS) Reference Architecture Mode* and hardware-enforced *Trusted Execution Environment (TEE)* security specifications are also included, paving the foundation for future *Open Policy Agent (OPA)* Policy-as-Code specifications.
3. **Regulatory & Standards WIP:** Mathematical equations, and resource usage models are compiled such as sustainability metrics and cost of ownership.

These frameworks are explicitly collected to ensure data sovereignty and process ownership at the factory perimeter.

### D. Analysis Method: Continuous Assurance

Data analysis follows a conventional theory-generating case study approach to categorize and re-sensitize the linear, siloed structures of semiconductor manufacturing into accessible, circular system dynamics design canvases.

The structural logic of the system is analyzed by synthesizing the document-driven requirements into System Dynamics Causal Loop Diagrams (CLDs). The structural logic of the system is analysed by synthesizing the document-driven requirements into System Dynamics Causal Loop Diagrams (CLDs). To conceptualize the macro feedback mechanisms governing these loops, this step adapts and expands the "Atom-to-Values" System of Systems (SoS) loops for AI compute [29], which extended J. Barber's Sustainable Production and Consumption (SPaC) framework.

By utilizing the oscillating, recycling, and balancing dynamics inherent in the "Atom-to-Values" paradigm, the analysis embeds a formal Triple Bottom Line (TBL) framework to orchestrate the competing decision weights across the three professional proxy agents (Profit, Planet, and People). The resulting framework enables the future embedding of quantitative formulas, metrology error tolerances, and impact thresholds directly into declarative Open Policy Agent (OPA) specifications.

These system dynamics feed directly into Multi-Criteria Decision Analysis (MCDA) formulations, demonstrating how multi-criteria trade-offs stabilize the socio-technical system and transform regulatory compliance from a retroactive operational burden into a real-time, proactive manufacturing capability, shaping manufacturing and business services [30].

## IV. Research Findings: The Professional Proxy

The proposed architecture operationalizes an automated SoS by embedding role-based professional proxy agents directly into the 6-Layer architecture, thereby leveraging trustworthy data for advanced sharing and modelling, enabling corporate governance automation while protecting manufacturing trade secrets and intellectual property (IP).

### *A. Threading Processes with Agentic Proxies*

Extending the 6-Layer Semiconductor SSbD architecture, three autonomous agentic threads connect introduce system dynamics for proactive data reuse, reporting, and modelling. These three threads, representing relevant functions of smart fabs (facility, process and finance), collaborate proactively to deliver operational intelligence that is sustainable (i.e., planet, people, and profits), as shown in Table I.

TABLE I. THREADING SMART SUSTAINABLE SSBD PROCESSES

| Stack Layer | 1. Facility Proxy 🌏 | 2. Process Proxy 👷 | 3. Finance Proxy 💰 |
|---|---|---|---|
| 🌐 *Layer 6: Resilient & Net-Zero Value Chain* | | | |
| 🏭 *Layer 5: Industry 4.0 Integration* | | | |
| 📏 *Layer 4: Metrology & Smart Manufacturing* | | | |
| 📁 *Layer 3: Federated Industrial Space* | | | |
| 📝 *Layer 2: RegTech & CBAM Compliance* | | | |
| 🌱 *Layer 1: Safe & Sustainable by Design (SSbD)* | | | |

*Note*: Cell content indicates active functional process (assignment) weight.

Based on division of labour function assignments and cross-functional collaboration needs, each proxy agent thread has at least three major operational layers:

1. **The Fab Facility Manager Proxy**: Supervises utility resource circularity, wastewater parameters, and chemical processing systems. It handles raw facility telemetry and environmental constraints at the physical-digital boundary, optimizing resource circularity without access to proprietary tool recipes or financial ledger rules.
2. **The Process Engineering Manager Proxy**: Acts as the custodian of the fabrication facility's core process IP. Covering mainly Layers 3, 4, and 5, it executes TEE enclave of sensitive data, multi-variant Virtual Metrology (VM) yield models, and machine learning adaptations, preventing high-value process parameters from leaking.
3. **The Procurement & Finance Accountant Proxy**: Operates at the finance and compliance interfaces. Covering mainly Layers 2, 3, and 6, it manages declarative Policy-as-Code ledger rules, outward IDS Connectors, and carbon-tariff or sustainability-linked financial mandate directives (e.g., CBAM, CSDDD) without exposing underlying tool telemetry.

Processes can follow a Mandate-Execute-Verify loop. For instance, it receives a mandate of "Verify Scope 3 emissions for Isopropyl alcohol (IPA) Batch #88 and corresponding solvent wastewater records", execute queries to gather both vendor and metrology data using the IDS connectors, and then verify the retrieved data points, thereby checking its compliance with related recycling and toxicity standards.

### *B. Walkthrough of Functional Threads*

Since energy-related GHG abatement efficiencies is needed in the near-term timeframe [IRDS™ ESHS ESSF roadmap], it is critical to expand the material flows between semiconductor and petrochemical sectors, such as the inclusion of IPA usage and waste. The paper proposes a conceptual GHG (emission) or EU SSbD (waste toxicity) models, complementing the existing water and energy models, to be implemented in the Layer 4 as predictive models that leverage the real-time monitoring and compliance data flows, that in turn fuels the machine learning and AI for digital twin innovations. Layer 3 acts as the cross-functional convergence plane for data provenance support.

#### *1) Environmental Stewardship & Facility Infrastructure*

The autonomous Fab Facility Manager Proxy thread enforces resource circularity and maximizes conservation targets, as defined by IEEE or SEMI F98 standards.

- **Layer 1 (SSbD) Intersection:** The proxy monitors physical telemetry from both procured chemical materials and cleanroom infrastructure components, tracking environment impact indicators such as fresh water and wastewater variants, modelling input-output models.
- **Layer 4 (Metrology) Intersection:** The proxy ingests these variables to construct the Total Available Conservation Opportunity (TACO) matrix, other water or resource measurement, and even advanced soft-sensor opportunities for water, energy, emission, or chemical toxicity modelling and management.
- **Layer 5 (Industry 4.0 Integration) Intersection:** Rather than dumping unparsed sensor logs into an external cloud, the proxy streams the validated matrix or models into the site's live Digital Twin. It runs neural network simulations to project how scaling up key KPIs, such as chemical mixture water reclaim ($Q_K$) under varying conditions.

Table I suggests this proxy should own the processes involving especially Layer 5 as the digital twin advocate, while covering also Layers 1 and 4.

#### *2) Processing Sensor Reliability & Data Sovereignty*

The Process Engineering Manager Proxy executes processing resource and time adjustments by deploying proactive production processing profiles that protect wafer yield and device performance.

- **Layer 3 (Federated Industrial Space) Intersection:** When an external query targets the processing profile of IPA Batch #88, Thread 2 isolates the execution inside a hardware-rooted TEE. It processes vendor-supplied chemical purity metadata alongside internal tool data, preventing proprietary recipe run-times or specific chemical dispense rates from leaking to external value chain competitors.
- **Layer 4 (Metrology) Intersection:** The proxy intercepts tool-level telemetry to assess the IRDS challenge metric: *Non-Beneficial Process Chemical Usage*. Because processing tools continuously purge solvent during idle states to prevent nozzle clogging, the proxy executes a localized predictive model. It calculates whether the idle purge rate of IPA Batch #88 can be dynamically shortened or modified without causing defects on the active wafer lot.
- **Layer 5 (Industry 4.0 Integration) Intersection:** Once the predictive adjustments are validated against the local digital twin thresholds, the proxy wraps the engineering facts into a privacy-preserving, cryptographically signed token. This token proves optimization execution while strictly cloaking underlying cleanroom configuration IP.

Table I suggests this proxy own the core execution of data validation, serving as the Layer 4 Predictive Model Advocate.

### 3) *RegTech & FinTech Accountability*

The Fab Procurement & Finance Accountant Proxy owns the compliance processes by designing and transforming technical proofs into globally verifiable economic assets.

- **Layer 2 (RegTech Compliance) Intersection:** This thread initiates the MEV lifecycle by translating macro corporate mandates, such as EU SSbD Per- and poly-fluoroalkyl substances (PFAS) controls or CBAM carbon intensity limits into machine-readable OPA schemas.
- **Layer 3 (Federated Industrial Space) Intersection:** Operating at the perimeter interface, the proxy manages an outward-facing IDS Connector to handle customs documentation and cross-border trade matching. It ingests the secured toxicity or emission sensor data from the TEE enclave directly to the connectors. For instance, emission data can be used to validate cross-border carbon tax credits, and verify import/export tariff compliance with customs authorities—all without exposing the fab's tool recipes.
- **Layer 6 (Resilient & Net-Zero Value Chain) Intersection:** The proxy passes these audited tax and tariff calculations into the IRDS Cost of Ownership **(**CoO) financial framework to compute the final net product-level liabilities and resource cost avoidance. It anchors these validated financial and engineering facts into an immutable DPP token exported to global OEMs, customs clearing networks, and green banking consortiums, unlocking automated sustainability-linked financing and automated tariff clearances based on trustworthy data provenance.

Table I suggests this proxy initiate the processes the compliance workflow and oversees the final ledger entries, serving as the Layer 2 RegTech & Compliance Advocate.

## C. *The SoS Agentic Relay Race: 5-Step Framework*

The IRDS™ 2024 roadmap Section 7.6 calls for a proactive data-driven framework to overcome rising operational complexity. The proposed architecture operationalizes a risk-mitigation sequence as a coordinated socio-technical relay race, replacing central controllers with automated data hand-offs and synchronization boundaries. Fig. 1 integrates not only the three proxies from Table I but also show how the "batons" are passed in the relay races in 5 consecutive steps.

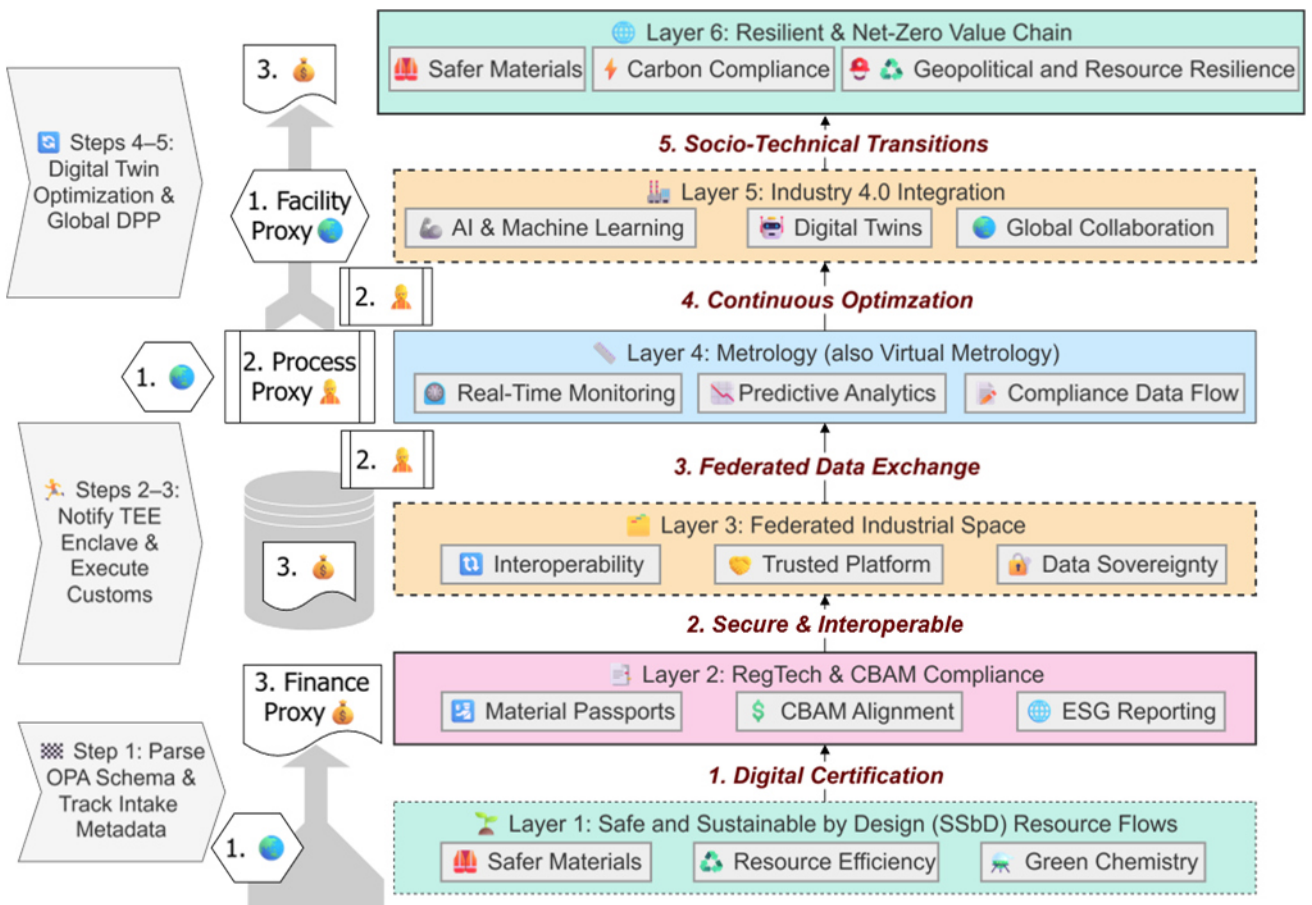

Fig. 1. SSbD Reference Architecture and Agentic Relay Race for the Semiconductor Industry

- **Step 1 (From Layer 1 to 2):** Responding to chemical footprint challenges, the Facility (Thread 1) and Finance (Thread 3) proxies collaborate to align the bottom lines of Planet and Profit. The lifecycle begins at Layer 1 where the Facility proxy captures real-time physical telemetry of incoming material lots (e.g., IPA Batch #88) and logs physical consumption variables. This matches the OPA constraints defined at Layer 2 by the Finance proxy, which translates macro-level corporate mandates, cross-border tax codes, and customs directives into machine-readable verification targets.
- **Steps 2-3 (From Layer 2 to 3 and 4):** Responding to processing compliance challenges, the system dynamics transition from administrative mandates to federated execution. The Finance proxy passes the "baton" across the IDS Connector at Layer 3, triggering a signal that notifies the Process proxy (Thread 2) of an active compliance tracking request. Thread 2 isolates the execution inside a hardware-rooted TEE enclave at Layer 3, cross-checking vendor-supplied chemical purity metadata against customs classifications and internal yield metrics. This triggers tool-level adjustments via Layer 4 metrology links, executing predictive models to shorten or route non-beneficial idle solvent purge flows directly into specialized waste segregation lines without exposing the fab's underlying process recipes.
- **Steps 4-5 (From Layer 4 to 5 and 6):** The technical proofs are synthesized into macro-economic value. At Layer 5, the soft-sensor metrology outputs from Layer 4 fuel the live, in-fab Digital Twin. The Facility proxy acts as the digital twin advocate here, running localized neural network simulations to verify that the optimized chemical recycling profiles do not exceed municipal wastewater toxicity thresholds or cause long-term filter membrane degradation. Once validated, this engineering fact is passed to Layer 6, where the Finance proxy integrates the data with the IRDS CoO financial framework. This automates the tracking of high-value byproducts—such as upcycling waste streams to save significant upstream smelting or manufacturing energy—and anchors the results into an immutable DPP token to secure automated cross-border customs clearance, CBAM carbon-tariff reconciliations, and sustainability-linked financing.

Altogether, from procuring and taking in materials, to recording trustworthy data that enables advanced water, energy, emission, and chemicals modeling and applications for in-fab digital twins and supply-chain-wide orchestration, this agentic relay race resolves the Data Sovereignty Paradox. It transforms local cleanroom manufacturing discipline into an unshakeable ecosystem of global regulatory and economic trust built entirely on absolute data provenance.

## D. *System Dynamics Validation & Operationalizing the "Process-as-Product" Framework*

To validate how this reference architecture resolves the core technological trade-offs inherent in modern semiconductor manufacturing—such as the interaction where maximizing water and chemical upcycling traditionally drives up local energy expenditures and brine generation—the architecture must be operationalized through a "process-as-product" governance fabric. This philosophy treats cross-layer data orchestration with the same engineering rigor and statistical process controls applied to physical wafers.

Instead of allowing factory subsystems to manage these trade-offs in isolated silos, the System of Systems (SoS) architecture routes these conflicting variables into the Layer 3 Multi-Criteria Decision Analysis (MCDA) convergence plane. To validate and stress-test this dynamic equilibrium under authentic manufacturing conditions, smart fabs can deploy focused, cross-functional pilot projects centred around a single high-impact asset class:

- **Targeted Asset Isolation:** A single high-capacity Ultra-Pure Water (UPW) recycling loop or an industrial chemical scrubbing and solvent segregation stack (e.g., handling IPA waste) is isolated as a bounded test domain.
- **Proxy Orchestration Activation:** Within this track, all three professional proxy agents are activated simultaneously. Thread 1 ingests sub-second physical sensor telemetry to drive the Layer 5 digital twin; Thread 2 manages inward TEE enclave evaluations to protect yield-critical recipes while executing predictive models at Layer 4; and Thread 3 runs declarative Policy-as-Code schemas at Layer 2 to generate external compliance and carbon-intensity tokens.
- **Dynamic Trade-Off Resolution:** When Thread 1 aggressively ramps up reclamation to hit sustainability targets, the system dynamics loop monitors the resulting power spikes and brine concentrations. If the energy increase triggers a Layer 2 policy alert for exceeding CBAM carbon intensity limits, the Layer 3 MCDA engine weighs the water/chemical savings against carbon-tariff exposure. Thread 2 coordinates predictive idle tool purges via smart fab communication links, matching real-time tool states with the reclamation loop's active capacity to minimize non-beneficial consumption without risking particle defects.
- **Human-in-the-Loop Integration:** Fab directors and engineers establish macro-strategic boundaries, performance weights, and exceptional override thresholds. Meanwhile, the autonomous proxies handle high-velocity data ingestion, enclave verification loops, and cross-border ledger transactions.

This targeted implementation validates the automated passing of compliance batons across the six layers. Treating the information lifecycle as a physical process establishes an ecosystem of trust built on absolute data provenance, providing the structural scaffolding necessary to scale verifiable safety and sustainability across fragmented global semiconductor value chains.

Together, these six layers form a coherent architecture where RegTech, Industry 5.0, and SSbD converge to build a safer, net-zero semiconductor value chain.

## V. DISCUSSION: RESOLVING THE PROVENANCE PROBLEM VIA AGENTIC RELAY RACES

### A. *Answering Research Question(s)*

The proposed six-layer architecture operationalizes risk mitigation not through centralized controllers, but as a coordinated socio-technical relay race that systematically resolves the Data Sovereignty Paradox. To demonstrate its structural logic and architectural viability, the framework synthesizes the answers to our three operationalized research questions into a unified macro-orchestration loop.

#### 1) *The Micro-Alignment Steps*

- **Operations & Autonomy (Steps 1 to 2):** To allow localized proxies to intercept tool telemetry without degrading cleanroom autonomy, the relay race begins at the physical-digital boundary. The facility proxy captures real-time physical variables of material intake at Layer 1. This payload is handed up to Layer 2, where the finance proxy parses it against machine-readable constraints. This aligns planetary boundaries and factory profit directly at the edge before any raw data is exposed.
- **Finance & Accountability (Steps 2–3):** To translate resource metrics into verifiable accounting proofs without leaking tool recipes, the finance proxy passes the operational "baton" across the industrial data space connector at Layer 3. This secure boundary hand-off triggers the process proxy to isolate execution inside a hardware-rooted TEE enclave. The proxy cross-checks chemical purity against customs classifications and uses Layer 4 metrology links to execute predictive models, modifying purge flows without disclosing underlying tool configurations.
- **Global Value Chain Harmonization (Steps 4–5):** Technical **proofs** are ultimately synthesized into macroeconomic value to secure a resilient, net-zero Industry 5.0 ecosystem. The process proxy routes soft-sensor outputs to Layer 5, where the facility proxy runs neural network simulations within the in-fab digital twin to optimize chemical recycling profiles. Once validated, this fact is passed to Layer 6, where the finance proxy integrates the data into financial frameworks, anchoring the results into an immutable DPP token for automated customs clearance, CBAM tariff reconciliations, and CSDDD auditing.

#### 2) *The Macro-Orchestration of Triple Bottom Line*

Weaving localized walkthroughs of our three functional threads prove that regulatory compliance is an engine for value creation.

- **Thread 1: Environmental Stewardship & Facility Infrastructure (The Planet Dimension):** This thread binds sustainability straight to cleanroom hardware across Layers 1, 4, and 5. By verifying localized solvent upcycling before waste leaves the fab boundary, it resolves the reporting "velocity gap" and validates safe chemical substitution in real time.
- **Thread 2: Processing Sensor Reliability & Data Sovereignty (The People Dimension):** Serving as the predictive model advocate across Layers 3, 4, and 5, this thread confines Virtual Metrology adaptations within secure enclaves. It protects the engineer's proprietary process configurations while dynamically stabilizing tool states to eliminate chemical waste and wafer defects.
- **Thread 3: RegTech & FinTech Accountability (The Profit Dimension):** This thread commercializes cleanroom discipline. Operating across Layers 2, 3, and 6, it maps engineering tokens to Harmonized System customs codes for organic chemicals, automating carbon-tariff reconciliations and unlocking automated, sustainability-linked financing based on trusted data provenance.

### B. Match and Contribution

This study directly aligns with the core research objectives by bridging engineering discipline with macroeconomic socio-technical transition Industry 5.0 orchestration:

- **Addressing Technology Management:** Rather than treating AI, digital twins, and federated data sharing as isolated novelties, this paper provides a structured approach to managing their integration. It introduces a zero-trust multi-agent framework that safely coordinates these emerging technologies within high-stakes, capital-intensive semiconductor manufacturing environments.
- **Providing Practical Frameworks:** The proposed six-layer architecture and five-step agentic relay race move beyond aspirational ESG goals. They offer industrial engineers, fab accountants, and supply chain managers a concrete, highly scalable reference blueprint to operationalize data hand-offs, enforce zero-trust security boundaries, and automate compliance audits.
- **Analysing Implementation Challenges:** This study explicitly tackles the core implementation bottlenecks modern fabs face: the reporting "velocity gap" between continuous cleanroom telemetry and static corporate disclosures, the threat of IP leakage during multi-stakeholder value-chain audits, and the friction of matching local manufacturing metrics with rigid cross-border customs classifications. We resolve these challenges by combining hardware-rooted TEE enclaves with policy-as-code OPA schemas
- **Focusing on Value Creation:** By transforming local regulatory compliance into a distinct competitive advantage, the framework redefines factory-floor constraints as economic opportunities. It automates the tracking of high-value chemical byproducts, prevents costly wafer yield defects through inline metrology corrections, and structures verifiable data proofs into immutable DPP tokens. This directly mitigates global carbon-tariff liabilities and unlocks automated, sustainability-linked credit facilities, establishing a highly resilient, net-zero business ecosystem for the modern industrial landscape.

## VI. CONCLUSION

### A. Limitations

To advance the concept from architectural blueprint to industrial deployment, future research must address:

- **Empirical Validation:** The proposed framework and its corresponding agentic roles represent a conceptual synthesis and have not yet undergone large-scale empirical validation or longitudinal testing within active fab production environments.
- **Maturity Alignment:** Cross-company integration of these proxies remains subject to the evolving maturity of international data spaces and decentralized compliance standards.
- **Governance Complexity:** Relying on autonomous agents as specialized proxies introduces unresolved challenges regarding accountability and AI governance across fragmented multi-tier vendor supply chains.

### B. Concluding Remarks

This study contributes to the progress of the field by introducing a zero-trust multi-agent architecture that moves the semiconductor sector from aspirational green goals to technically verifiable outcomes. By replacing centralized controllers with automated data hand-offs and synchronization boundaries, this research provides a unified path to resolve the Data Sovereignty Paradox.

The primary insight gained is that compliance can be embedded directly into the digital bedrock of design and production. This transforms regulatory mandates from operational burdens into catalysts for ecosystem-wide innovation. By operationalizing specialized agentic roles—from environmental edge tracking to strategic policy execution—the reference framework aligns localized industrial metrology with macro global governance. This establishes the foundational trust framework necessary for capital-intensive ecosystems to drive an industrial transition that is simultaneously intelligent, sustainable, and geopolitically resilient.

### C. Future Work and Perspectives

To advance these socio-technical dynamics, future research should investigate the integration of MCDA formulations directly with declarative LLM agentic skill files (e.g., skills.md). This framework allows professional proxies to autonomously evaluate multi-criteria trade-offs—such as balancing rigorous wafer yield constraints against immediate environmental footprints or cross-border labor compliance. Codifying these complex decision matrices into structured, readable markdown skill profiles enables autonomous agents to dynamically adapt their operational logic. This capability directly reshapes the traditional MNC considerations of talent globalization, specifically within Global Business Services (GBS) and Global Mobility Services (GMS), where shifting international mandates highlight critical concerns regarding labor rights and immigration policies [31], [32], [33]. By embedding localized regulatory constraints directly into the agents' operational profiles, this methodology bridges the gap between senior human expertise and automated execution across global manufacturing and mobility corridors.

Future research and standardization efforts should prioritize the formation of multipolar industrial coalitions that bridge the diverse standard-setting landscapes identified in our framework. Specifically, there is an urgent need to harmonize the technical industrial standards originating from Taiwan (SEMI) and the USA (NIST) with the rigorous socio-technical and environmental mandates of the EU (JRC, ECHA) and international bodies like the ISO and ITU. This effort must focus on codifying "Agentic Skills" into interoperable, open-source protocols, allowing professional proxies to maintain continuity across jurisdictions and eliminating the translation bottlenecks that currently fragment the global semiconductor value chain.

In this evolving landscape, the European Semiconductor Manufacturing Company (ESMC) occupies a unique strategic position to pilot these architectural innovations. By leveraging the "Supply Online 360" platform—TSMC's digital bedrock for responsible supply chain management—ESMC can serve as the primary global laboratory for the six-layer SSbD architecture. The platform's existing infrastructure for digitized sustainability modules, its Supplier Sustainability Academy, and its dedicated worker grievance channels provide a mature environment for integrating autonomous agentic proxies.

Implementing these proxies within the Dresden cluster offers a rare opportunity to bridge the "Copy Exactly!" manufacturing ethos with the "Comply Exactly" regulatory mandates of CSDDD and CBAM. This integration will demonstrate that the transition to Industry 5.0 is not merely a regional compliance exercise but a scalable, competitive blueprint for resilient and net-zero semiconductor manufacturing worldwide.

## Acknowledgment

The authors gratefully acknowledge the constructive feedback of the double-blind peer reviewers whose scrutiny strengthened this work; all remaining errors and interpretations are solely the authors' own responsibility.